%% file: conf-0438.tex
\documentclass[aps,prl,preprint,tightenlines,superscriptaddress,showpacs,byrevtex]{revtex4}
\usepackage{epsfig}
\usepackage{dcolumn}  

\graphicspath{{ps}}

\begin{document}



\preprint{\vbox{ \hbox{   }
                 \hbox{BELLE-CONF-0438}
                 \hbox{ICHEP04 8-0864} 
}}

\title{
 \quad\\[0.5cm]
Studies of CP violation in $B^0 \rightarrow J/\psi K^{*0}$ decays}

\input author-conf2004

\noaffiliation

\begin{abstract}
An angular analysis for $B$ decays into two vector mesons 
gives parameters sensitive to new physics.
They are the decay amplitudes of the three helicity states, the asymmetry 
in triple product, and the CP violating phase. The measurements of
these quantities are performed for $B^0 \rightarrow J/\psi K^{*0}$
decays in a data sample with 253 fb$^{-1}$ taken by the Belle detector 
in the KEKB asymmetric $e^+e^-$ collider. 
With the time-integrated angular analysis, the decay
amplitudes' moduli and phase angles are measured to be $|A_0|^2=0.585\pm 0.012\pm 0.09$,
$|A_{\parallel}|^2=0.233\pm 0.013\pm 0.008$,
$|A_\perp|^2=0.181\pm 0.012\pm 0.008$,
$\arg(A_\parallel)=2.888\pm 0.090\pm 0.008$ radians, and
$\arg(A_\perp)=0.239\pm 0.064\pm 0.010$ radians. 
The difference between
$\arg(A_\parallel)$ and $\arg(A_\perp)$ is $2.649\pm0.110$ radians, which is
shifted from $\pi$ by more than $4\sigma$; this can be interpreted as
evidence for the existence of the final state interaction.
The amplitude values are used in triple product correlations to obtain
asymmetries of
$A_T^{(1)}=0.101\pm 0.033\pm 0.007$ and $A_T^{(2)}=-0.102\pm 0.032\pm 0.003$
for $B^0$ meson decays, and 
$\bar{A}_T^{(1)}=0.056\pm 0.030\pm 0.006$ and $\bar{A}_T^{(2)} =
-0.091\pm 0.028\pm 0.003$ for $\overline{B^0}$ meson decays. 
The time dependent angular analysis gives the 
measurements of CP violating parameters $\sin 2\phi_1$ and $\cos 2\phi_1$
to be $0.30 \pm 0.32 \pm 0.02$ and $-0.31 \pm 0.91 \pm 0.10$,
respectively. The value of $\cos 2\phi_1$ is $-0.56 \pm 0.86 \pm
0.11$ if $\sin 2\phi_1$ is fixed at the world average value (0.731).
\end{abstract}


\maketitle

\tighten

{\renewcommand{\thefootnote}{\fnsymbol{footnote}}}
\setcounter{footnote}{0}

\section{Introduction}
An angular analysis of $B$ meson decay to two vector mesons can be a
sensitive probe of new physics. Since the analysis 
is equivalent to the
simultaneous study of three decay modes corresponding to the possible
helicity states of two vector mesons, the effect of the new physics
that might appear in the interference among the states can be observed
cleanly with the cancellation of most systematic effects. 

Recently, the anomaly in the decay amplitudes in $B \rightarrow \phi
K^{*}$ has been reported where the longitudinal amplitude is 
lower than the Standard Model prediction\cite{phiKstar}. For
the precise study of such anomalies, a comparison with
theoretically clean 
reference modes is necessary. The decay $B^0\rightarrow J/\psi K^{*0}$ 
is an ideal mode for this purpose since it is a tree dominated
mode with a very low penguin contribution where the theoretical
ambiguity on the CP-sensitive parameter $\sin 2\phi_1$\cite{Hadronic} 
is less than 1\% level\cite{jpsitheory}.

There are three classes of parameters that can be used to probe new
physics obtained through the angular analysis. The first is the 
measurement of the decay amplitudes of the three helicity states. 
They can be obtained by the time-integrated
angular analysis to flavor specific decays. 
The comparison of the amplitudes between flavor-separated samples
probes the direct CP violation. 
The second is the triple product correlation, which can be 
extracted from the measured decay amplitudes.
This quantity is sensitive to the T-violation. 
The third class is comprised of the
CP parameters ($\sin 2\phi_1$ and $\cos 2\phi_1$) 
that are measured
through the time-dependent angular analysis. In particular, the
measurement of $\cos 2\phi_1$, which appears in the time-dependent
interference terms, is important
both to solve the two-fold ambiguity in $2\phi_1$ and to test the
consistency of this determination with the more precise value from
other $b \to c\bar{c}s$ decays.

In this paper, we report the measurements of all three classes of 
parameters for $B^0 \rightarrow J/\psi K^{*0}$ decays. 

\section{Data sample}

The data sample used in this analysis corresponds to an integrated
luminosity of 253 fb$^{-1}$ recorded with the Belle detector\cite{Belle}
at the KEKB electron-positron collider\cite{KEKB}.
Events are required to satisfy the hadronic event selection 
criteria\cite{Hadronic} and have $R_2<0.5$,
where $R_2$ is the ratio of the second to zeroth Fox-Wolfram event
shape moments\cite{fw}.

The reconstruction of $J/\psi$ candidates is performed using the dilepton
decays $J/\psi \rightarrow e^+e^-$ and $\mu^+\mu^-$. For
the $e^+e^-$ mode,
a track is identified as an electron or positron by a comparison of
the energy
measured in the electromagnetic
calorimeter (ECL) with the momentum measured in the central
drift chamber (CDC), the shape of the cluster energy deposit in the
ECL, the specific ionization ($dE/dx$) measured in the CDC, and 
the light yield in the
aerogel \v{C}erenkov counters (ACC).
The likelihoods that the track is an electron or a hadron are
determined from these measurements. The likelihood ratio is required
to be consistent with the electron hypothesis.
To correct for energy lost by final state radiation,  the energy of
any cluster in the ECL within 0.05 radians of the initial track momentum is added
to that of the track.
The invariant mass of each so-augmented electron-positron pair 
is calculated, and the pair is associated with $J/\psi \rightarrow
e^+e^-$ if the
mass is in the range $2.95~{\rm GeV}/c^2 < M(e^+e^-) < 3.15~{\rm
GeV}/c^2$.
Tracks are identified as muons rather than hadrons
by means of likelihoods based on
$(i)$ a comparison of the number of layers with associated hits in
the muon detector (KLM) with the  number expected based on momentum and
$(ii)$ the energy of the associated hit in the ECL.
An oppositely charged pair of muons  is identified as arising from
$J/\psi \rightarrow \mu^+\mu^-$ if the invariant mass is in the range
$3.05~{\rm GeV}/c^2 < M(\mu^+\mu^-) < 3.15~{\rm GeV}/c^2$.
To improve the momentum resolution, a kinematic fit that uses the
$J/\psi$ mass as a constraint
is performed on $J/\psi$ candidates passing the above selections.

Candidate $K^{*0}$ mesons
are reconstructed using the decay modes
$K^{*0}\rightarrow K^+\pi^-$ and $K^{*0}\rightarrow K_S^0\pi^0$.
A track is identified as a kaon based on the ratio of its likelihoods
to be a kaon {\it vs} a pion.
These likelihoods are
obtained from the measurements mentioned above from the TOF, CDC and ACC
detectors. 
Tracks that are not identified
as kaons and not used as leptons in the $J/\psi$ reconstruction
are treated as charged pion candidates.
$K_S^0$ candidates are
reconstructed from pairs of oppositely charged tracks that satisfy
three conditions: 1) the
distance of closest approach of each
track to the nominal interaction point is larger than 0.03 cm,
2) the angle between the $K_S^0$ momentum vector and the vector
displacement of the $K_S^0$ vertex point from the $J/\psi$ vertex is
less than 0.15 radians, and
3) the reconstructed decay vertex of the $K_S^0$ is at least 0.1 cm
away from the interaction point.
Each pair with  an invariant mass satisfying
$0.47~{\rm GeV}/c^2 < M(\pi^+\pi^-) < 0.52~
{\rm GeV}/c^2$ is identified as a $K_S^0$ meson.
To improve the mass resolution, the pion momenta are re-fitted,
constraining both tracks to originate at the reconstructed $K_S^0$ vertex.

Candidate $\pi^0$ mesons are reconstructed from
clusters in the ECL
that are unmatched to charged tracks and having energy greater than
$40~{\rm MeV}$.
A photon pair with an invariant mass
in the range
$0.125~{\rm GeV}/c^2 < M(\gamma\gamma) < 0.145~{\rm GeV}/c^2$ is identified
as a $\pi^0$. A mass-constrained fit is performed to obtain the
momentum of the $\pi^0$ meson.
A $K \pi$ pair is considered a $K^*$ candidate if the invariant
mass of the pair is within 74 MeV/$c^2$ of the nominal $K^*(892)$ mass.

Candidate $B^0$ mesons are reconstructed by selecting
events with a $J/\psi$ and a $K^*$ meson
and examining two
quantities in the center-of-mass of the $\Upsilon$(4S): the
beam-constrained mass ($M_{\rm bc}$)
and the energy difference between the $B$
candidate and the beam energy ($\Delta E$).
The beam-constrained mass, which is the invariant mass of a reconstructed
$J/\psi$ and $K^*$ calculated taking the energy to be the beam
energy, is required to be
in the range $5.27-5.29~{\rm GeV}/c^2$.
For the mode with $K^{*0}\rightarrow K^+\pi-$,
$|\Delta E|$ is required to be less than $30~{\rm MeV}$, while
$\Delta E$ is required to satisfy $-50~{\rm MeV} <
\Delta E < 30~{\rm MeV}$ for $K^{*0}\rightarrow K_S^0\pi^0$.

\begin{figure}
\centerline{\mbox{\psfig{figure=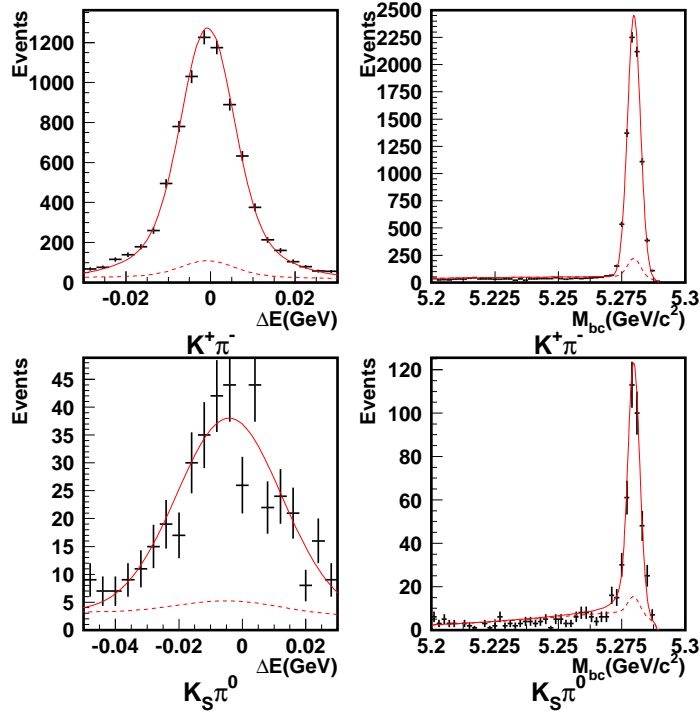,width=10cm}}}
\caption{$\Delta E$ and $M_{\rm bc}$ distributions for reconstructed $B^0\rightarrow
J/\psi K^{*0}(K^+\pi^-)$ (upper figures) and $B^0\rightarrow
J/\psi K^{*0}(K_S^0\pi^0)$(lower figures) samples. Solid lines show
the results of two dimensional fits while dashed lines are
estimated background contaminations.}
\label{dEMbcplot}
\end{figure}

To eliminate slow $\pi^0$ backgrounds,
the angle $\theta_{K^*}$ of the kaon with respect
to the $K^*$ direction in the  $K^*$ rest frame is required
to satisfy ${\rm cos}\theta_{K^*}<0.8$. This is equivalent to 
demanding that
the $\pi^0$ momentum be greater than $175~{\rm MeV}/c$.
When an event contains more than one candidate passing the above requirements,
the combination for which both $M_{\rm bc}$ and $\Delta E$ are the closest to
the $B^0$ mass and zero, respectively, is selected.
After all the selections, {\bf 8194}
$K^{*0}\rightarrow{K^+\pi^-}$ candidates and
{\bf 354} $K^{*0}\rightarrow K_S^0\pi^0$ candidates remain.
Figure ~\ref{dEMbcplot} shows the distributions of $M_{\rm bc}$ and
$\Delta E$ for the samples.

\section{Measurement of decay amplitudes}

\begin{figure}
\centerline{\mbox{\psfig{figure=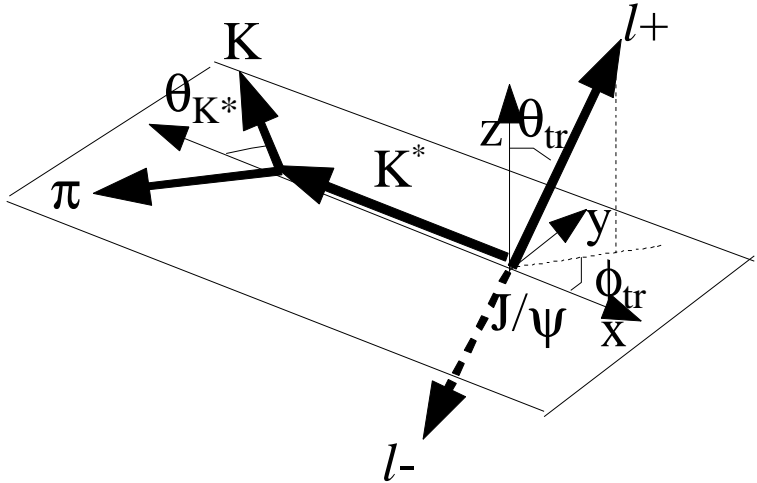,width=10cm}}}
\caption{The definition of transversity angles.}
\label{fig:angle-defs}
\end{figure}

The decay amplitudes of $B\rightarrow J/\psi K^*$ decays are
measured in the transversity basis\cite{transversity}.
The definition of the angles is shown in Fig.~\ref{fig:angle-defs}.
The direction of motion of the $J/\psi$ in the rest frame of the $B$
candidate is defined to be the $x$-axis. The $y$-axis is chosen along
the direction of the projection of the $K$ momentum into the plane
perpendicular to the $x$-axis in the $B$ rest frame. The $x$-$y$ plane
contains the momenta of the $J/\psi$, the $K$, and the $\pi$ mesons.
The $z$-axis is then perpendicular to the $x$-$y$ plane
according to the right-hand rule.
The angle between the the positive lepton ($l^+$) daughter and the
$z$-axis 
in the $J/\psi$ rest frame is defined as $\theta_{tr}$.
The angle between the $x$-axis and the projection
of the $l^+$ momentum
onto the $x$-$y$ plane is defined as $\phi_{tr}$ in the same frame. 
The angle $\theta_{K^*}$ is defined as described in the previous section.

The distribution of these three
angles for $B \rightarrow J/\psi  K^*$ decays
is described in terms of three amplitudes\cite{Yamamoto}:
\begin{eqnarray}
\frac{1}{\Gamma}\frac{d\Gamma}{d\cos\theta_{tr}d\cos\theta_{K^*}d\phi_{tr}}
& = &    \frac{9}{32\pi}\times
     [ 2\cos^2\theta_{K^*}(1-\sin^2\theta_{tr}\cos^2\phi_{tr})|A_0|^2
\nonumber \\
& &   + \sin^2\theta_{K^*}(1-\sin^2\theta_{tr}\sin^2\phi_{tr})|A_{\parallel}|^2
\nonumber \\
& &   + \sin^2\theta_{K^*}\sin^2\theta_{tr}|A_{\perp}|^2 \nonumber \\
& &   + \eta \sin^2\theta_{K^*}\sin 2\theta_{tr} \sin\phi_{tr} 
Im(A_{\parallel}^*A_{\perp}) \nonumber \\
& &   - \frac{1}{\sqrt{2}} \sin 2\theta_{K^*} \sin^2\theta_{tr} \sin 
2\phi_{tr} Re(A_0^*A_{\parallel}) \nonumber \\
& &   + \eta \frac{1}{\sqrt{2}} \sin 2\theta_{K^*} \sin 2\theta_{tr}
\cos\phi_{tr} Im(A_0^*A_{\perp}) ], \label{theory}
\end{eqnarray}
where $A_0$, $A_{\parallel}$ and $A_{\perp}$ are the complex
amplitudes of the three helicity states in the trasversity basis, and
$\eta = +1~(-1)$ for $B^0$ ($\overline{B^0}$). The coefficient
$|A_0|^2$ denotes the longitudinal polarization of $J/\psi$ while
$|A_{\perp}|^2$ ($|A_{\parallel}|^2)$ gives the transverse polarization
component along the $z$-axis ($y$-axis). In the decay mode where
$K^{*0}\rightarrow K_S^0  \pi^0$, $|A_0|^2 + |A_{\parallel}|^2$
($|A_{\perp}|^2$) is the component corresponding to the CP-even (CP-odd) state.

The amplitudes are determined by fitting this function to the
measured three-dimensional distribution in $\theta_{tr}$, $\phi_{tr}$
and $\theta_{K^*}$, taking into account the detection efficiency and 
background.
The resolution of the angular measurements is estimated by Monte Carlo 
(MC) simulation
and found to be typically less than 0.02 radians.
The value of $\eta$ is determined from the charge of pions
or kaons used in the $K^*$ reconstruction.
We use only $B^0\rightarrow J/\psi K^{*0}(K^+ \pi^-)$ decays for the
measurement since $\eta$ is not well-defined in $B^0\rightarrow
J/\psi K^{*0}(K_S^0 \pi^0)$ decays.

The fit is performed using an unbinned maximum likelihood
method. The probability density function (${\mathcal G}$) is defined using the
theoretical distribution in (\ref{theory}) and can be expressed as
\begin{eqnarray}
{\mathcal G} (x,y,z, M_{\rm bc},\Delta E) & = & N \times
  [ f_{sig}(M_{\rm bc},\Delta E) \times \epsilon(x,y,z) \times \frac{1}{\Gamma}
\frac{d^3\Gamma}{dxdydz}(x,y,z) \nonumber \\
& & +~ \sum_i f_{cf}^i(M_{\rm bc},\Delta E) \times {\mathcal A}_{cf}(x,y,z)\nonumber \\
& & +~ f_{nr}(M_{\rm bc},\Delta E) \times {\mathcal A}_{nr}(x,y,z) \nonumber \\
& & +~ f_{cb}(M_{\rm bc},\Delta E) \times {\mathcal A}_{cb}(x,y,z) ], \label{PDF1}
\end{eqnarray}
where $x = {\rm cos}\theta_{tr}$, $y = \phi_{tr}$, and $z = {\rm
cos}\theta_{K^*}$, $N$ is the normalization factor of ${\mathcal G}$,
$\epsilon(x,y,z)$ is the detection efficiency as a function of the three 
angles, and
${\mathcal A}_{cf}$, ${\mathcal A}_{nr}$, ${\mathcal A}_{cb}$ 
are the angular shapes for the
cross-feed, non-resonant and combinatorial backgrounds, respectively.

Here, $f_{sig}(M_{\rm bc},\Delta E)$, $f_{cf}^i(M_{\rm bc},\Delta E)$, 
$f_{nr}(M_{\rm bc},\Delta E)$, and
$f_{cb}(M_{\rm bc},\Delta E)$ are the fractions of
signal events, cross-feed contamination, non-resonant
contamination and combinatorial background, respectively, as a function of 
$M_{\rm bc}$ and $\Delta E$.
These fractions are obtained with a data subsample of $140\,{\rm
fb}^{-1}$ by the procedure
given in Ref.\cite{angleana-29} with an improvement 
to use both $M_{\rm bc}$ and $\Delta E$ in the fits.

The detection efficiency function $\epsilon(x,y,z)$ is
obtained from a large MC sample of 6 million events
generated polarized with the decay amplitudes set at the values of
the previous measurement\cite{angleana-29}.
Events are histogrammed in a $20\times20\times20$ grid in the
${\rm cos}\theta_{tr}-\phi_{tr}-{\rm cos}\theta_{K^*}$
cube. The distribution is fitted to the product of three
one-dimensional polynomials.

The angular distribution function for the cross-feed background
(${\mathcal A}_{cf}$) is
determined from MC simulation.
The function for the non-resonant production (${\mathcal A}_{nr}$) is
determined from
events in the sideband of the $K\pi$ mass distribution ($1.0\, {\rm GeV}/c^2 <
M(K\pi) < 1.3\, {\rm GeV}/c^2$).
The distribution for the combinatorial background (${\mathcal A}_{cb}$) is
obtained from events in the sideband with $5.2\, {\rm GeV}/c^2 
< M_{\rm bc} < 5.26\, {\rm GeV}/c^2$.
These normalized distributions are parameterized as the product of
three one-dimensional polynomials whose
parameters are determined from the fit.

\begin{figure}
\centerline{\mbox{\psfig{figure=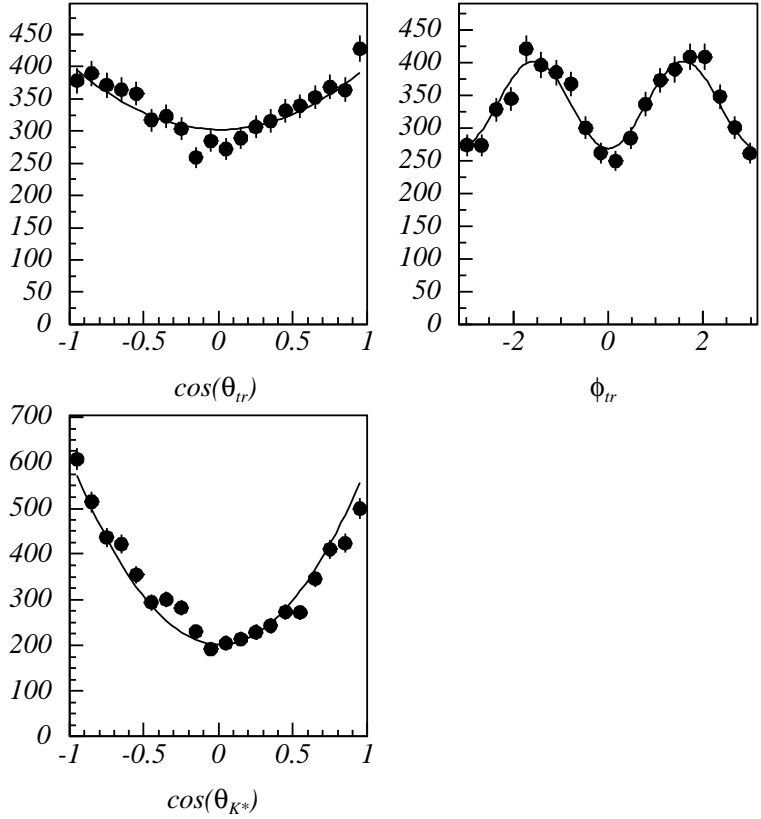,width=10cm}}}
\caption{Distributions of projected angles. Solid lines show results
of the fit. The data points are corrected for the detector efficiency
and the backgrounds are subtracted.}
\label{fig:angfinal}
\end{figure}

In the fit,
the phase of $A_0$ is defined to be zero
relative to those of the other amplitudes since the
overall phase of the decay amplitudes is arbitrary. The other five
parameters, $|A_0|^2$, $|A_{\parallel}|^2$, $|A_{\perp}|^2$,
$\arg(A_{\parallel})$ and $\arg(A_{\perp})$ are free parameters in the fit.
The normalization condition of the amplitudes
\begin{equation}
|A_0|^2 + |A_{\parallel}|^2 + |A_{\perp}|^2 = 1 \label{NORM}
\end{equation}
is taken into account by adopting the extended likelihood defined as
\begin{equation}
-\ln L = -\sum_{i=1}^{N_{obs}}\ln {\mathcal G}_i + N_{exp} - N_{obs}\ln(N_{exp})
\end{equation}
where
$N_{obs}$ is the number of events used for the fit. 
$N_{exp}$ is defined to be $N_{obs} \cdot (|A_0|^2 + |A_{\parallel}|^2 +
|A_{\perp}|^2)$ to incorporate the normalization condition. 
The normalization $N$ of ${\mathcal G}$ is recalculated by numerical 
integration 
whenever the fit parameters change.
The parameter values determined from the fit are summarized in
Table~\ref{table:amp-all} and
the projected angular distributions are shown in
Fig.~\ref{fig:angfinal}. The distributions are corrected for the effects
of detector acceptance and backgrounds.
Small discrepancies from 0 or $\pi$ are observed in
$\arg(A_{\parallel})$ and $\arg(A_{\perp})$. The difference of these two 
phases is $2.649\pm0.110$ radians which is shifted from $\pi$ by
more than $4 \sigma$. This is read to be evidence for the existence
of final
state interaction.

There are two choices for the imaginary phases of the helicity
amplitudes\cite{Mahiko}. We obtain the above values by taking
the conventional choice of the phases noted as solution I 
in Ref.\cite{Mahiko}.
If the alternative set of the phases is used, the imaginary phases become
$\arg(A_{\parallel}) \rightarrow - \arg(A_{\parallel})$, and
$\arg(A_{\perp}) \rightarrow \pi - \arg(A_{\perp})$.

The parameter values are also measured for $B^0$ (4121 candidates)
and $\overline{B^0}$ (4073 candidates)
samples separately and the results are given in
Table~\ref{table:amp-all}.
The measured values are consistent between both $B^0$ and $\overline{B^0}$
decays, indicating no evidence for direct CP violation.
\begin{table}
\caption{Summary of measured decay amplitudes.}
\begin{center}
\begin{tabular}{|c|ccc|}
\hline
 & Flavor averaged & $B^0$ & $\overline{B^0}$ \\ \hline
$|A_0|^2$           & $0.585\pm0.012$ & $0.581\pm0.016$ & $0.589\pm0.016$ \\
$|A_{\parallel}|^2$ & $0.233\pm0.013$ & $0.220\pm0.018$ & $0.246\pm0.018$ \\
$|A_{\perp}|^2$     & $0.181\pm0.012$ & $0.199\pm0.018$ & $0.164\pm0.017$ \\
$\arg(A_{\parallel})$  & $2.888\pm0.090$ & $2.937\pm0.136$ & $2.854\pm0.120$ \\
$\arg(A_{\perp})$      & $0.239\pm0.064$ & $0.303\pm0.090$ & $0.182\pm0.089$ \\ 
\hline
\end{tabular}
\label{table:amp-all}
\end{center}
\end{table}

Systematic uncertainties in the fit are determined as follows:
1) efficiency function (MC statistics and effect of polarization), 2)
angular distribution functions for backgrounds, 3) background
fractions, 4) slow pion efficiency, and 5) polarization effect in
non-resonant decays.
These contributions to the systematic error are
summarized in Table~\ref{table:syse-angle}. The dominant contribution
arises from the uncertainty in the detection efficiency
for slow pions. This is estimated by comparing 
the results with a cut at $\cos \theta_{K^*} <0.9$. 
The angular shape in non-resonant backgrounds is a potential source of
uncertainty since it might contain tails of other resonances whose 
polarization is different. The
uncertainty is determined by comparing with the results obtained assuming the
angular distribution to be that of phase space decay.
\begin{table}
\begin{center}
\caption{Systematic errors in the measurement of decay amplitudes.
  \label{angsyse}}
\begin{tabular}{|c|ccccc|}
\hline
Item & $|A_0|^2$ & $|A_\parallel|^2$ & $|A_\perp|^2$ & $\arg(A_{\parallel})$ &
$\arg({A_\perp})$ \\ \hline
Efficiency           & 0.003 & 0.002 & 0.001 & 0.001 & 0.003 \\
PDF for backgrounds  & 0.002 & 0.002 & 0.002 & 0.004 & 0.006 \\
Background fractions & 0.001 & 0.002 & 0.003 & 0.001 & 0.002 \\
Slow pion efficiency & 0.008 & 0.007 & 0.007 & 0.006 & 0.007 \\
Polarization in NR   & 0.002 & 0.002 & 0.003 & 0.003 & 0.003 \\
\hline
Total                & 0.009 & 0.008 & 0.008 & 0.008 & 0.010 \\
\hline
\end{tabular}
\label{table:syse-angle}
\end{center}
\end{table}

\section{Triple Product Correlations}

The T-violating triple product correlations are sensitive probes to
new physics effects\cite{TP}. The asymmetry of the triple product is defined as
\begin{equation}
A_T = \frac{\Gamma(\vec{v_1}\cdot(\vec{v_2}\times\vec{v_3}) > 0 ) - 
	\Gamma(\vec{v_1}\cdot(\vec{v_2}\times\vec{v_3}) < 0 )}
	{\Gamma(\vec{v_1}\cdot(\vec{v_2}\times\vec{v_3}) > 0 ) + 
	\Gamma(\vec{v_1}\cdot(\vec{v_2}\times\vec{v_3}) < 0 )}
\end{equation}
where $\vec{v_1}$ is the momentum of $J/\psi$ or $K^{*0}$ meson, while
$\vec{v_2}$ and $\vec{v_3}$ are the polarization vectors of $J/\psi$
and $K^{*0}$ meson, respectively. Experimentally, two asymmetries can
be defined using the measured decay amplitudes for $B^0$ decays, as follows:
\begin{equation}
A_T^{(1)} = \frac{Im(A_{\perp}A_{0}^*)}{A_0^2+A_{\parallel}^2+A_{\perp}^2} \\
\;\;\;\;
A_T^{(2)} = \frac{Im(A_{\perp}A_{\parallel}^*)}{A_0^2+A_{\parallel}^2+A_{\perp}^2}.
\label{eq:TP}
\end{equation}
The corresponding asymmetries for $\overline{B^0}$ decays are
defined as $\bar{A}_T^{(1)}$ and $\bar{A}_T^{(2)}$. The Standard 
Model predicts tiny values for these asymmetries and no difference
between $B^0$ and $\overline{B^0}$ mesons.

By putting the measured amplitude values in Eq.~\ref{eq:TP}, the
triple product asymmetries are obtained as
\begin{eqnarray}
	A_T^{(1)} & = & 0.101 \pm 0.033 \pm 0.007 \nonumber \\
	A_T^{(2)} & = & -0.102 \pm 0.032 \pm 0.003 \nonumber \\
	{\bar A}_T^{(1)} & = & 0.056 \pm 0.030 \pm 0.006  \nonumber \\
	{\bar A}_T^{(2)} & = & -0.091 \pm 0.028 \pm 0.003  \nonumber
\end{eqnarray}
As seen, all the obtained triple product asymmetries are close to zero, furthermore, no
difference between $A_T$'s and $\bar{A}_T$'s is observed.

\section{Measurement of CP violation parameters}

The CP violation parameters $\sin 2\phi_1$ and $\cos
2\phi_1$ are measured by studying the angular distributions in the CP
decay mode $B^0 \rightarrow J/\psi K^{*0};K^{*0} \rightarrow K_S^0\pi^0$ 
as a function of the decay time difference between $B^0$ and
$\overline{B^0}$ ($\Delta t$). The $\Delta t$-dependent form of 
Eq.~\ref{theory} becomes\cite{Yamamoto}:
\begin{equation}
\frac{d^4\Gamma(\theta_{tr},\phi_{tr},\theta_{K^*},\Delta t)}
{d\cos\theta_{tr} d\phi_{tr} d\cos\theta_{K^*} d\Delta t} 
= \frac{e^{-|\Delta t|/\tau_{B^0}}}{2\tau_{B^0}} \sum_{i=1}^{6}
g_i (\theta_{tr},\phi_{tr},\theta_{K^*}) a_i(\Delta t) 
\label{tadf}
\end{equation}
where $\tau_B$ is the lifetime of a $B^0$ meson, and the angular terms 
are expressed as:
\begin{eqnarray}
g_1 & = & 2\cos^2\theta_{K^*}(1-\sin^2\theta_{tr}\cos^2\phi_{tr}) \\
g_2 & = & \sin^2\theta_{K^*}(1-\sin^2\theta_{tr}\sin^2\phi_{tr}) \\
g_3 & = & \sin^2\theta_{K^*}sin^2\theta_{tr} \\
g_4 & = & \frac{-1}{\sqrt{2}}\sin
2\theta_{K^*}\sin^2\theta_{tr}\sin2\phi_{tr} \\
g_5 & = & \sin^2\theta_{K^*}\sin2\theta_{tr}\sin\phi_{tr} \\
g_6 & = & \frac{1}{\sqrt{2}}\sin
2\theta_{K^*}\sin2\theta_{tr}\cos\phi_{tr} 
\end{eqnarray}
while the amplitude terms are expressed as:
\begin{eqnarray}
a_1 & = & |A_0|^2(1+\eta\sin2\phi_1 \sin\Delta m\Delta t) \\
a_2 & = & |A_{\parallel}|^2(1+\eta\sin2\phi_1 \sin\Delta m\Delta t) \\
a_3 & = & |A_{\perp}|^2(1-\eta\sin2\phi_1 \sin\Delta m\Delta t) \\
a_4 & = & Re(A_{\parallel}^*A_0)(1+\eta \sin2\phi_1 \sin\Delta
m\Delta t) \\
a_5 & = & \eta Im(A_{\parallel}^*A_{\perp})\cos \Delta m\Delta t - 
\eta Re(A_{\parallel}^*A_{\perp})\cos 2\phi_1 \sin \Delta m \Delta t \\
a_6 & = & \eta Im(A_0^*A_{\perp})\cos \Delta m\Delta t - 
\eta Re(A_0^*A_{\perp})\cos 2\phi_1 \sin \Delta m \Delta t. 
\end{eqnarray}
Here, $\Delta m$ is a $B^0-\overline{B^0}$ mixing parameter.
$\eta$ is $+1$ for $B^0$ and $-1$ for $\overline{B^0}$.
$A_0$, $A_{\parallel}$ and $A_{\perp}$ are the decay amplitudes 
given in the previous section.
Two CP violation parameters appear in the formula,
{\it viz}, $\sin2\phi_1$ and $\cos2\phi_1$. 
The latter parameter appears in the interference terms $a_5$ and
$a_6$.

The procedures to measure the decay time difference and to determine
the flavor of decaying $B^0$ meson are described
elsewhere\cite{Hadronic}. 
The values of $\sin2\phi_1$ and $\cos2\phi_1$ are determined by
fitting the function to the measured angles and $\Delta t$,
 taking into account the detection efficiency and background.
The fit is done using an unbinned maximum likelihood method. The
probability density function for an event is defined as
\begin{eqnarray}
{\mathcal P} & = & 
f_{sig}(M_{\rm bc},\Delta E)
\epsilon(\theta_{tr},\phi_{tr},\theta_{K^*}) 
\frac{d^4\Gamma(\theta_{tr},\phi_{tr},\theta_{K^*},\Delta t)}
{d\cos\theta_{tr} d\phi_{tr} d\cos\theta_{K^*} d\Delta t}  \nonumber \\
& & + \frac{e^{-|\Delta t|/\tau_{B^0}}}{2\tau_{B^0}} \{
  \sum_i f_{cf}^i(M_{\rm bc},\Delta E)
{\mathcal A}_{cf}(\theta_{tr},\phi_{tr},\theta_{K^*})  \nonumber \\
& & ~~~~~~~~~~~~~~~~ + f_{nr}(M_{\rm bc},\Delta E) 
{\mathcal A}_{nr}(\theta_{tr},\phi_{tr},\theta_{K^*}) \}
\nonumber \\
& & + \delta(\Delta t) f_{cb}(M_{\rm bc},\Delta E) 
{\mathcal A}_{cb}(\theta_{tr},\phi_{tr},\theta_{K^*}). 
\label{eqn:timepdf}
\end{eqnarray}
Here, $f_{sig}$, $f_{cf}$, $f_{nr}$ and $f_{cb}$ are the fractions of 
signal, cross feed, non-resonant production, and combinatorial
background components, respectively,
while
${\mathcal A}_{cf}$, ${\mathcal A}_{nr}$ and ${\mathcal A}_{cb}$ are
the corresponding three-dimensional angular shape functions. 
Also, $\epsilon$ is a three
dimensional detection efficiency function for the signal. 
These functions are the same as those used in the decay amplitude
measurement. 
The flavor tagging procedure gives
the flavor $q$ of tag-side $B$ meson and the probability $w$ that this 
flavor determination is incorrect.
To account for the effect of wrong tagging in this fit, 
$\eta$ is replaced with $-q(1-2w)$ within the signal angular
distribution in Eq.~\ref{eqn:timepdf}.

Each term in ${\mathcal P}$ is then convolved with the appropriate resolution
functions separately for the signal, backgrounds having the
$B^0$ lifetime (namely, cross feeds and non-resonant production) 
and the combinatorial background with a zero-lifetime $\delta$-function shape, 
which takes into account the effect of vertex position smearing caused by 
the detector resolution, charmed meson decays, and poorly
reconstructed tracks. The
resolution parameters are calculated event by event using the 
energy of the reconstructed $B$ meson, the direction of the $B$ meson with respect 
to the beam axis and the momentum of the $B$ meson in the center-of-mass
frame~\cite{Blife}.

In the fit, the helicity amplitudes are fixed at the values determined 
in the previous section.
The values of the lifetime and mixing
parameter are set at the values given in Ref.~\cite{PDG}.
The CP parameters $\sin 2\phi_1$ and $\cos 2\phi_1$ are the only free
parameters in the fit.
From the fit to the data, we obtain
\begin{eqnarray}
	\sin 2\phi_1 & = & 0.30 \pm 0.32 \pm 0.02, \nonumber \\
	\cos 2\phi_1 & = & -0.31 \pm 0.90 \pm 0.10. \nonumber
\end{eqnarray}
When we fix the value of $\sin 2\phi_1$ at the world average value
(0.731)\cite{HFAG}, the value of $\cos 2\phi_1$ becomes
\begin{equation}
	\cos 2\phi_1 = -0.56 \pm 0.86 \pm 0.11 \nonumber
\end{equation}
We obtain the above values by taking a set of the phases given in the 
previous section. If the
alternative set of the phases is used in the fit,
the sign of $\cos2\phi_1$ flips, giving
$\cos 2\phi_1 = +0.31 \pm 0.86 \pm 0.11$, while $\sin
2\phi_1$ does not change.

Systematic uncertainties in the fit are determined in the same manner as 
those in $b \to c\bar{c}s$ $\sin 2\phi_1$ measurement\cite{Hadronic}. 
In addition, the
uncertainties that come from the angular analysis (decay amplitudes,
background angular shapes, and background fractions) are estimated.
The uncertainty in decay amplitudes are determined by varying
the values by one standard deviation. The uncertainty in the background fractions
and angular shapes are estimated in the same manner as that in the
decay amplitude measurement. 

\begin{figure}
\centerline{\mbox{\psfig{figure=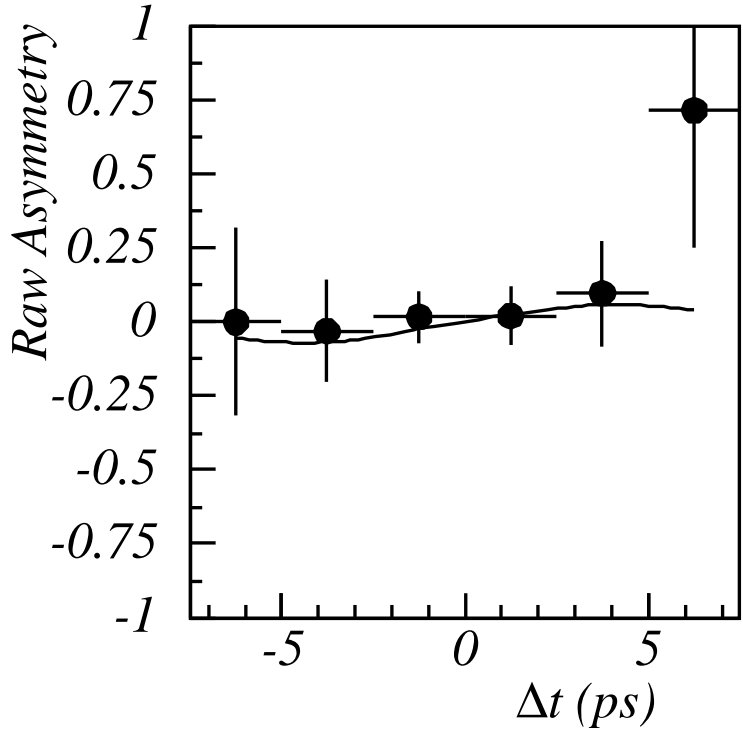,width=10cm}}}
\caption{Raw asymmetry in measured $\Delta t$ between samples tagged
as $B^0$ and $\overline{B^0}$. The solid 
line shows the projection of the fit.}
\label{fig:asym}
\end{figure}

\begin{table}
\begin{center}
\caption{Systematic errors in the measured CP parameters.
\label{table:sysecp}}
\begin{tabular}{|c|cc|c|}
\hline
Item                        & $\sin 2\phi_1$ & $\cos 2\phi_1$ & $\cos
2\phi_1$ ($\sin 2\phi_1 = 0.731$) \\ \hline 
Vertexing                   & 0.008 & 0.050 & 0.051 \\
Resolution parameters       & 0.005 & 0.010 & 0.010 \\
Wrong tagging fractions     & 0.009 & 0.061 & 0.062 \\
$\delta$-function for BG     & 0.006 & 0.004 & 0.005 \\
$\tau_B$ and $\Delta m$     & 0.004 & 0.010 & 0.010 \\
Decay amplitudes            & 0.005 & 0.041 & 0.048 \\
Angular shape for backgrounds   & 0.007 & 0.025 & 0.044 \\
Background fraction         & 0.011 & 0.021 & 0.021 \\ \hline
Total                       & 0.020 & 0.096 & 0.107 \\
\hline
\end{tabular}
\end{center}
\end{table}
The raw asymmetry in the measured $\Delta t$ between samples tagged as
$B^0$ and $\overline{B^0}$ is shown in Fig.~\ref{fig:asym} together
with the result of the fit.

The systematics in the fit are checked by applying the same fitting
procedure to the sample of $B\rightarrow J/\psi K^{*0}(K^+\pi^-)$
decays. We obtain following values for parameters $\sin
2\phi_1$ and $\cos 2\phi_1$:
\begin{eqnarray}
``\sin 2\phi_1``  & = & -0.037 \pm 0.068 \nonumber \\
``\cos 2\phi_1``  & = & 0.103 \pm  0.179 \nonumber
\end{eqnarray}
These are consistent with zero, as expected.

\section{Conclusion}

A full angular analysis is performed 
for $B^0\rightarrow J/\psi K^{*0}$ ($K^{*0} \rightarrow K^+\pi^-$)
decays. The complex decay amplitudes of
three helicity final states are measured by the
simultaneous fit to three transversity angles and the best fit values are
$|A_0|^2  =  0.585 \pm 0.012 \pm 0.009$,
$|A_{\parallel}|^2 = 0.233 \pm 0.013 \pm 0.008$,
$|A_\perp|^2  =  0.181 \pm 0.012 \pm 0.008$,
$\arg(A_{\parallel})  =  2.888 \pm 0.090 \pm 0.008$ radians, and
$\arg(A_{\perp}) = 0.239 \pm 0.064 \pm 0.010$ radians,
where $\arg(A_0)$ is defined to be zero.
There are two choices for the set of imaginary phases.
We obtain the above values by taking the conventional choice of phases.
The measured value of $|A_{\perp}|^2$ shows that the CP even component
dominates in $B^0 \rightarrow J/\psi K^{*0}; K^{*0} \rightarrow K_S^0
\pi^0$ decays. Small discrepancies from $\pi$ or $0$ are observed in
$\arg(A_{\parallel})$ and $\arg(A_{\perp})$, respectively. The difference of these two 
phases is calculated to be $2.649\pm0.110$ radians and the shift from $\pi$ is 
more than $4 \sigma$. This is read to be an evidence for the existence
of final state interactions.
The amplitudes are also measured for $B^0$ and
$\overline{B^0}$ decays separately. The best fit values are consistent 
between them and no direct CP violating effect is observed.

The measured amplitudes are converted into the asymmetries 
in the triple product correlations: the best fit values are
$A_T^{(1)} = 0.101 \pm 0.033 \pm 0.007$ and
$A_T^{(2)} = -0.102 \pm 0.032 \pm 0.003$ for $B^0$ mesons, and 
$\bar{A}_T^{(1)} = 0.056 \pm 0.030 \pm 0.006$ and 
$\bar{A}_T^{(2)} = -0.091 \pm 0.028 \pm 0.003$ for $\overline{B^0}$ mesons. 
The differences between two asymmetries for $B^0$ and $\overline{B^0}$
mesons are consistent with zero and no new physics
effects are observed.

The time dependent angular analysis is performed for 
$B^0\rightarrow J/\psi K^{*0}$ ($K^{*0} \rightarrow K_S^0\pi^0$) decays.
By the simultaneous fit to the 
three transversity  angles and $\Delta t$, the CP violation parameters 
$\sin 2\phi_1$ and $\cos 2\phi_1$ are determined to be
$\sin 2\phi_1 = 0.30 \pm 0.32 \pm 0.02$ and 
$\cos 2\phi_1 = -0.31 \pm 0.91 \pm 0.10$. 
Fixing $\sin 2\phi_1$ at the world average value (0.73) gives 
$\cos 2\phi_1 = -0.56 \pm 0.86 \pm 0.11$. 
The decay amplitudes are fixed at the values measured in the time
integrated fit with the conventional choice of imaginary phases.
When taking the other choice, the
sign of $\cos 2\phi_1$ flips; however, 
we cannot put a constraint on the sign of $\cos 2\phi_1$
regardless of the choice of the imaginary phase set.

Table ~\ref{vs-babar} shows the comparison with previous measurements.
The measured decay amplitudes are consistent among all measurements. 
\begin{table}
\begin{center}
\caption{Comparison of obtained decay amplitudes and CP violation
 parameters with previous measurements. Here, the sign of $\cos 2\phi_1$ 
 of Belle measurements is put to be consistent with BaBar's choice of
 phases.}
\label{vs-babar}
\begin{tabular}{|c|c|c|c|}
\hline
& This Measurement & Belle\cite{angleana-29}\cite{ICHEP02-result} & BaBar\cite{BaBar-Moriond04} \\ \hline
Luminosity &   253${\rm fb}^{-1}$  & 29${\rm fb}^{-1}$ (78 for CP)&
83${\rm fb}^{-1}$ (113 for CP) \\ \hline
$|A_0|^2$            & $0.585 \pm 0.012 \pm 0.009$ & $0.62 \pm 0.02 \pm 0.03$  &
$0.566 \pm 0.012 \pm  0.005$     \\
$|A_{\parallel}|^2$  & $0.233 \pm 0.013 \pm 0.008$ &  - &
$0.204 \pm 0.015 \pm 0.005$         \\
$|A_{\perp}|^2$      & $0.181 \pm 0.012 \pm 0.008$ & $0.19 \pm 0.02 \pm 0.03$ &  
$0.230 \pm 0.015 \pm 0.004$    \\
$\arg(A_{\parallel})$ & $2.888 \pm 0.090 \pm 0.008$ & $2.83 \pm 0.19 \pm 0.08$ &
$2.729 \pm 0.101 \pm 0.052$        \\
$\arg(A_{\perp})$     & $0.239 \pm 0.064 \pm 0.010$ & $-0.09 \pm 0.13 \pm 0.06$ &
$0.184 \pm 0.070 \pm 0.046$   \\ \hline
$\sin 2\phi_1$       &   $0.30 \pm 0.32 \pm 0.02$   & $0.13 \pm 0.51 \pm 0.06$ & 
$-0.10 \pm 0.57$   \\
$\cos 2\phi_1$       &   $+0.31 \pm 0.91 \pm 0.11$  & $+1.40 \pm 1.28 \pm 0.19$ &
$+3.32^{+0.76}_{-0.96}\pm0.27$   \\ \hline
\end{tabular}
\end{center}
\end{table}

We thank the KEKB group for the excellent operation of the
accelerator, the KEK Cryogenics group for the efficient
operation of the solenoid, and the KEK computer group and
the National Institute of Informatics for valuable computing
and Super-SINET network support. We acknowledge support from
the Ministry of Education, Culture, Sports, Science, and
Technology of Japan and the Japan Society for the Promotion
of Science; the Australian Research Council and the
Australian Department of Education, Science and Training;
the National Science Foundation of China under contract
No.~10175071; the Department of Science and Technology of
India; the BK21 program of the Ministry of Education of
Korea and the CHEP SRC program of the Korea Science and
Engineering Foundation; the Polish State Committee for
Scientific Research under contract No.~2P03B 01324; the
Ministry of Science and Technology of the Russian
Federation; the Ministry of Education, Science and Sport of
the Republic of Slovenia; the National Science Council and
the Ministry of Education of Taiwan; and the U.S.
Department of Energy.

\end{document}

%% file: author-conf2004.tex
\affiliation{Aomori University, Aomori}
\affiliation{Budker Institute of Nuclear Physics, Novosibirsk}
\affiliation{Chiba University, Chiba}
\affiliation{Chonnam National University, Kwangju}
\affiliation{Chuo University, Tokyo}
\affiliation{University of Cincinnati, Cincinnati, Ohio 45221}
\affiliation{University of Frankfurt, Frankfurt}
\affiliation{Gyeongsang National University, Chinju}
\affiliation{University of Hawaii, Honolulu, Hawaii 96822}
\affiliation{High Energy Accelerator Research Organization (KEK), Tsukuba}
\affiliation{Hiroshima Institute of Technology, Hiroshima}
\affiliation{Institute of High Energy Physics, Chinese Academy of Sciences, Beijing}
\affiliation{Institute of High Energy Physics, Vienna}
\affiliation{Institute for Theoretical and Experimental Physics, Moscow}
\affiliation{J. Stefan Institute, Ljubljana}
\affiliation{Kanagawa University, Yokohama}
\affiliation{Korea University, Seoul}
\affiliation{Kyoto University, Kyoto}
\affiliation{Kyungpook National University, Taegu}
\affiliation{Swiss Federal Institute of Technology of Lausanne, EPFL, Lausanne}
\affiliation{University of Ljubljana, Ljubljana}
\affiliation{University of Maribor, Maribor}
\affiliation{University of Melbourne, Victoria}
\affiliation{Nagoya University, Nagoya}
\affiliation{Nara Women's University, Nara}
\affiliation{National Central University, Chung-li}
\affiliation{National Kaohsiung Normal University, Kaohsiung}
\affiliation{National United University, Miao Li}
\affiliation{Department of Physics, National Taiwan University, Taipei}
\affiliation{H. Niewodniczanski Institute of Nuclear Physics, Krakow}
\affiliation{Nihon Dental College, Niigata}
\affiliation{Niigata University, Niigata}
\affiliation{Osaka City University, Osaka}
\affiliation{Osaka University, Osaka}
\affiliation{Panjab University, Chandigarh}
\affiliation{Peking University, Beijing}
\affiliation{Princeton University, Princeton, New Jersey 08545}
\affiliation{RIKEN BNL Research Center, Upton, New York 11973}
\affiliation{Saga University, Saga}
\affiliation{University of Science and Technology of China, Hefei}
\affiliation{Seoul National University, Seoul}
\affiliation{Sungkyunkwan University, Suwon}
\affiliation{University of Sydney, Sydney NSW}
\affiliation{Tata Institute of Fundamental Research, Bombay}
\affiliation{Toho University, Funabashi}
\affiliation{Tohoku Gakuin University, Tagajo}
\affiliation{Tohoku University, Sendai}
\affiliation{Department of Physics, University of Tokyo, Tokyo}
\affiliation{Tokyo Institute of Technology, Tokyo}
\affiliation{Tokyo Metropolitan University, Tokyo}
\affiliation{Tokyo University of Agriculture and Technology, Tokyo}
\affiliation{Toyama National College of Maritime Technology, Toyama}
\affiliation{University of Tsukuba, Tsukuba}
\affiliation{Utkal University, Bhubaneswer}
\affiliation{Virginia Polytechnic Institute and State University, Blacksburg, Virginia 24061}
\affiliation{Yonsei University, Seoul}
  \author{K.~Abe}\affiliation{High Energy Accelerator Research Organization (KEK), Tsukuba} 
  \author{K.~Abe}\affiliation{Tohoku Gakuin University, Tagajo} 
  \author{N.~Abe}\affiliation{Tokyo Institute of Technology, Tokyo} 
  \author{I.~Adachi}\affiliation{High Energy Accelerator Research Organization (KEK), Tsukuba} 
  \author{H.~Aihara}\affiliation{Department of Physics, University of Tokyo, Tokyo} 
  \author{M.~Akatsu}\affiliation{Nagoya University, Nagoya} 
  \author{Y.~Asano}\affiliation{University of Tsukuba, Tsukuba} 
  \author{T.~Aso}\affiliation{Toyama National College of Maritime Technology, Toyama} 
  \author{V.~Aulchenko}\affiliation{Budker Institute of Nuclear Physics, Novosibirsk} 
  \author{T.~Aushev}\affiliation{Institute for Theoretical and Experimental Physics, Moscow} 
% ...
\collaboration{The Belle Collaboration}